\documentclass[twocolumn,british,english,american,twocolumn,superscriptaddress,showpacs,amsmath,amssymb,aps]{revtex4-1}
\usepackage[T1]{fontenc}
\usepackage[latin9]{inputenc}
\usepackage{babel}
\usepackage{verbatim}
\usepackage{textcomp}
\usepackage{amstext}
\usepackage{amssymb}
\usepackage{graphicx}
\usepackage{esint}
\usepackage{subscript}
\usepackage[unicode=true,pdfusetitle,
 bookmarks=true,bookmarksnumbered=false,bookmarksopen=false,
 breaklinks=false,pdfborder={0 0 0},backref=false,colorlinks=false]
 {hyperref}

\makeatletter

\newcommand{\lyxmathsym}[1]{\ifmmode\begingroup\def\b@ld{bold}
  \text{\ifx\math@version\b@ld\bfseries\fi#1}\endgroup\else#1\fi}

\usepackage{hyperref}
\hypersetup{colorlinks=true,citecolor=blue}
\usepackage{natbib}
\makeatother

\begin{document}
\selectlanguage{british}%

\preprint{APS/123-QED}

\title{Purcell-enhanced single-photon emission from nitrogen-vacancy centers
coupled to a tunable microcavity}

\author{Hanno Kaupp}

\affiliation{Fakultät für Physik, Ludwig-Maximilians-Universität, Schellingstraße~4,
80799~München, Germany}

\affiliation{Max-Planck-Institut für Quantenoptik, Hans-Kopfermann-Str.~1, 85748~Garching,
Germany}

\author{Thomas Hümmer}

\affiliation{Fakultät für Physik, Ludwig-Maximilians-Universität, Schellingstraße~4,
80799~München, Germany}

\affiliation{Max-Planck-Institut für Quantenoptik, Hans-Kopfermann-Str.~1, 85748~Garching,
Germany}

\author{Matthias Mader}

\affiliation{Fakultät für Physik, Ludwig-Maximilians-Universität, Schellingstraße~4,
80799~München, Germany}

\affiliation{Max-Planck-Institut für Quantenoptik, Hans-Kopfermann-Str.~1, 85748~Garching,
Germany}

\author{Benedikt Schlederer}

\affiliation{Fakultät für Physik, Ludwig-Maximilians-Universität, Schellingstraße~4,
80799~München, Germany}

\author{Julia Benedikter}

\affiliation{Fakultät für Physik, Ludwig-Maximilians-Universität, Schellingstraße~4,
80799~München, Germany}

\affiliation{Max-Planck-Institut für Quantenoptik, Hans-Kopfermann-Str.~1, 85748~Garching,
Germany}

\author{Philip Haeusser}

\affiliation{Fakultät für Physik, Ludwig-Maximilians-Universität, Schellingstraße~4,
80799~München, Germany}

\author{Huan-Cheng Chang}

\affiliation{Institute of Atomic and Molecular Sciences, Academia Sinica, Taipei
106, Taiwan}

\author{Helmut Fedder}

\affiliation{3. Physikalisches Institut, Universität Stuttgart, Pfaffenwaldring
57, 70569 Stuttgart, Germany}

\author{Theodor W. Hänsch}

\affiliation{Fakultät für Physik, Ludwig-Maximilians-Universität, Schellingstraße~4,
80799~München, Germany}

\affiliation{Max-Planck-Institut für Quantenoptik, Hans-Kopfermann-Str.~1, 85748~Garching,
Germany}

\author{David Hunger}

\email[]{david.hunger@physik.lmu.de}

\selectlanguage{british}%

\affiliation{Fakultät für Physik, Ludwig-Maximilians-Universität, Schellingstraße~4,
80799~München, Germany}

\affiliation{Max-Planck-Institut für Quantenoptik, Hans-Kopfermann-Str.~1, 85748~Garching,
Germany}

\date{\today}
\selectlanguage{american}%
\begin{abstract}
Optical microcavities are a powerful tool to enhance spontaneous emission
of individual quantum emitters. However, the broad emission spectra
encountered in the solid state at room temperature limit the influence
of a cavity, and call for ultra-small mode volume. We demonstrate
Purcell-enhanced single photon emission from nitrogen-vacancy (NV)
centers in nanodiamonds coupled to a tunable fiber-based microcavity
with a mode volume down to $1.0\,\lambda^{3}$. We record cavity-enhanced
fluorescence images and study several single emitters with one cavity.
The Purcell effect is evidenced by enhanced fluorescence collection,
as well as tunable fluorescence lifetime modification, and we infer
an effective Purcell factor of up to $2.0$. With numerical simulations,
we furthermore show that a novel regime for light confinement can
be achieved, where a Fabry-Perot mode is combined with additional
mode confinement by the nanocrystal itself. In this regime, effective
Purcell factors of up to 11 for NV centers and 63 for silicon vacancy
centers are feasible, holding promise for bright single photon sources
and efficient spin readout under ambient conditions.
\end{abstract}
\maketitle

\section{Introduction}

Solid-state-based quantum emitters such as the NV center in diamond
are promising for efficient single photon sources \citep{Kurtsiefer2000,Aharonovich2011},
quantum memories \citep{Neumann2010,Fuchs2011,Maurer2012}, and quantum
sensors \citep{Balasubramanian2008,Maze2008,Maletinsky2012}, with
functionality preserved also under ambient conditions. One of the
central challenges is to efficiently access the quantum properties
of the emitter by optical means. Free space coupling only achieves
low collection efficiency, and poses fundamental limits on achievable
fidelities and signal to noise ratios. Optical microcavities \citep{Vahala2003}
allow to enhance the light-matter interaction, and offer an increase
in spontaneous emission by the Purcell factor $C_{\textrm{eff}}=\frac{3(\lambda/n)^{3}}{4\pi^{2}}\frac{Q_{\textrm{eff}}}{V}$,
together with the potential for near-unity collection efficiency $\beta=C_{\textrm{eff}}/(C_{\textrm{eff}}+1)$.
Here, $V$ is the mode volume of the cavity, $n$ the refractive index,
and \foreignlanguage{english}{$Q_{\mathrm{eff}}=(Q_{c}^{-1}+Q_{em}^{-1})^{-1}$}
is the effective quality factor combining the quality factor of the
cavity ($Q_{\textrm{c}}$) and of the emitter ($Q_{\textrm{em}}$)
\citep{Auffeves2010,Meldrum2010}. For broadband emitters such as
the NV center, \foreignlanguage{english}{$Q_{\textrm{em}}$} as estimated
from the linewidth of the emission spectrum is small at room temperature
and limits the influence of high $Q_{\textrm{c}}$, and cavities with
ultra-small mode volume are required. A lot of effort has been put
into realizing coupled systems both with bulk micro- and nanocavities
\citep{Wolters2010,Englund2010,Barclay2011,Faraon2011,Sar2011,Hausmann2013,Lee2014,Li2015b,Riedrich-Moeller2015,Schroeder2016},
as well as tunable open-access microcavities \citep{Albrecht2013,Kaupp2013,Albrecht2014,Johnson2015}.
Although cavity-induced lifetime changes have been demonstrated, this
has been mostly limited to cryogenic temperatures however, where the
zero phonon line is narrow, and the extracted integrated count rates
remained well below spectrally integrated free space values. Furthermore,
the required smallest mode volumes have only been achieved in bulk
cavities, where limited spectral tunability as well as the difficulty
to find or place a single emitter in the cavity field maximum result
in low device yield.

Here, we use an open-access tunable microcavity \citep{Hunger2010,Toninelli2010,Dolan2010,Trupke2005,Barbour2011,Kelkar2015}
with ultra-small mode volume realized by advanced laser machining
to demonstrate Purcell-enhanced emission of NV centers in nanodiamonds.
We record cavity-enhanced fluorescence images of large areas of the
sample by scanning cavity microscopy \citep{Mader2015,Kelkar2015,Toninelli2010,Greuter2014}
and study several single emitters with one and the same cavity. Photon
collection rates from single NV centers of up to $1.6\times10^{6}$\,s$^{-1}$
are observed. Furthermore, we demonstrate continuous spontaneous emission
control by varying the mirror separation down to sub-$\mu$m mirror
separations, where we observe a maximal Purcell factor of $2.0$.
These results indicate that field confinement by the nanocrystal itself
is present in addition to the mode defined by the mirrors. For an
optimized geometry, numerical simulations predict an effective Purcell
factor of up to 11 for the broad room-temperature emission spectrum
of the NV center. At the same time, the outcoupling efficiency from
the cavity into a low-NA mode is as high as 51\%, limited by the intrinsic
loss of the used mirror coating.

\begin{figure}
\includegraphics[width=0.4\textwidth]{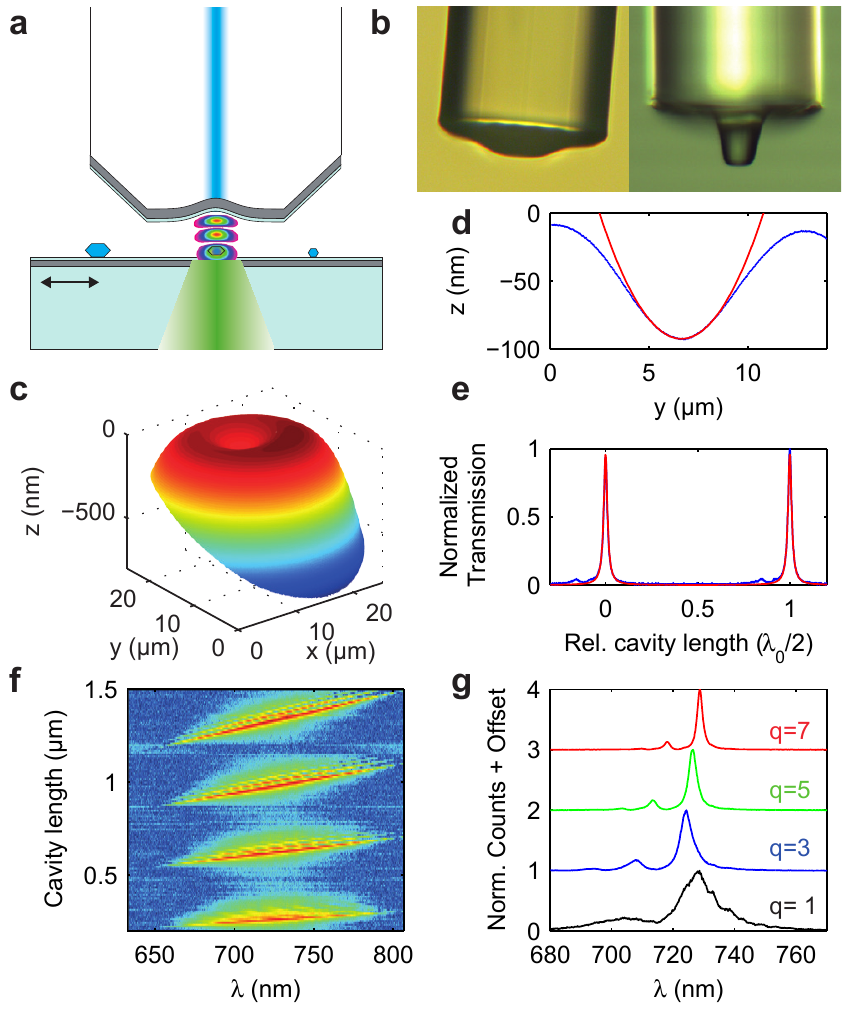}

\protect\caption{(a) Sketch of the cavity consisting of a laser-machined and mirror-coated
fiber and a macroscopic mirror carrying nano diamonds with NV centers.
The large mirror is mounted on a three-axis nanopositioning stage
for spatial scanning, the cavity length is controlled by an additional
piezo actuator. (b) Microscope images (20x) of two laser-shaped fiber
tips. (c) 3D profile of a laser-machined and silver-coated fiber tip.
(d) Cut through the center of the structure shown in (c) (blue) together
with a parabolic fit (red). (e) Cavity transmission probed with a
narrow-band laser as a function of the cavity length. (f) Series of
cavity transmission spectra under broadband illumination as a function
of the cavity length, showing tunability for cavity lengths down to
$\lambda_{0}/2$. (g) Individual spectra from (f) for different mode
orders $q$.\label{fig:1_Introduction}}
\end{figure}

\section{Ultra-small mode volume tunable cavity}

The microcavity is assembled of a planar mirror onto which the sample
is applied, and a concave mirror on the tip of an optical fiber \citep{Hunger2010},
see Fig.~\ref{fig:1_Introduction}(a). The tip of the cavity fiber
is shaped by advanced CO\textsubscript{2} laser machining \citep{Hunger2012,Ott2016}.
In a first step, the extent of the endfacet is tapered to enable sub-micron
mirror separations, which is crucial when aiming at smallest mode
volumes. Therefore, multiple laser pulses are applied in circular
patterns to the edge of the endfacet to crop the outer part of the
fiber, resulting in a protruding plateau with a diameter of typically
below $20\,\text{\textmu m}$. Figure~\ref{fig:1_Introduction}(b)
shows two examples of different shape. Next, a concave depression
aligned with the fiber core is produced within the center of the plateau
using a few weak laser pulses \footnote{For some geometries it is advantageous to create the concave profile
before tip shaping.}. To achieve smallest mode volumes, a compromise has to be chosen
between the size of the radius of curvature $r_{c}$ and the smallest
achievable mirror separation, since a small $r_{c}$ requires a deep
profile. Since we want to study the regime where the diamond crystal
provides additional field confinement (see below), we prioritize small
mirror separation and compromise on the radius of curvature. Figure~\ref{fig:1_Introduction}(c)
shows the topography of a machined fiber tip measured with white light
interferometry. The central part of the profile is well fitted by
a parabola, yielding a radius of curvature $r_{c}=90\:\text{\textmu m}$,
while the full profile fits to a Gaussian with $1/e$ diameter $D=7\,\text{\textmu m}$
and a structure depth $z<100\,\textrm{nm}$ (Fig.~\ref{fig:1_Introduction}(d)).

When choosing an optimal mirror coating, both the mirror reflectivity
and field penetration need to be considered. While dielectric coatings
provide highest reflectivity and low loss, they exhibit significant
field penetration into the coating, which increases the mode volume.
Optimizing the Purcell factor for broadband emitters shows that metal
coatings with few tens of nm penetration are advantageous, despite
the higher loss \citep{Steiner2006,Steiner2007,Kelkar2015}. We coat
the fiber tip with $60\,\textrm{nm}$ and the planar mirror with $33\,\textrm{nm}$
silver, both finished with a $20\,\textrm{nm}$ glass capping to prevent
oxidation. This yields reflectivities of $96\%$ and $88\%$ and absorption
loss of $4\%$, while scattering loss is found to be negligible. The
two mirrors define an open and tunable plano-concave Fabry-Perot cavity
that outcouples up to 51\% of the light through the planar mirror
into the detection channel (see Appendix). We measure a cavity finesse
of $\mathcal{F}=42\pm1$ at $\lambda_{0}=690$\,nm by recording the
cavity transmission of a narrow-band laser when tuning the cavity
length over one free spectral range (Fig.~\ref{fig:1_Introduction}(e)).

\begin{figure}
\includegraphics{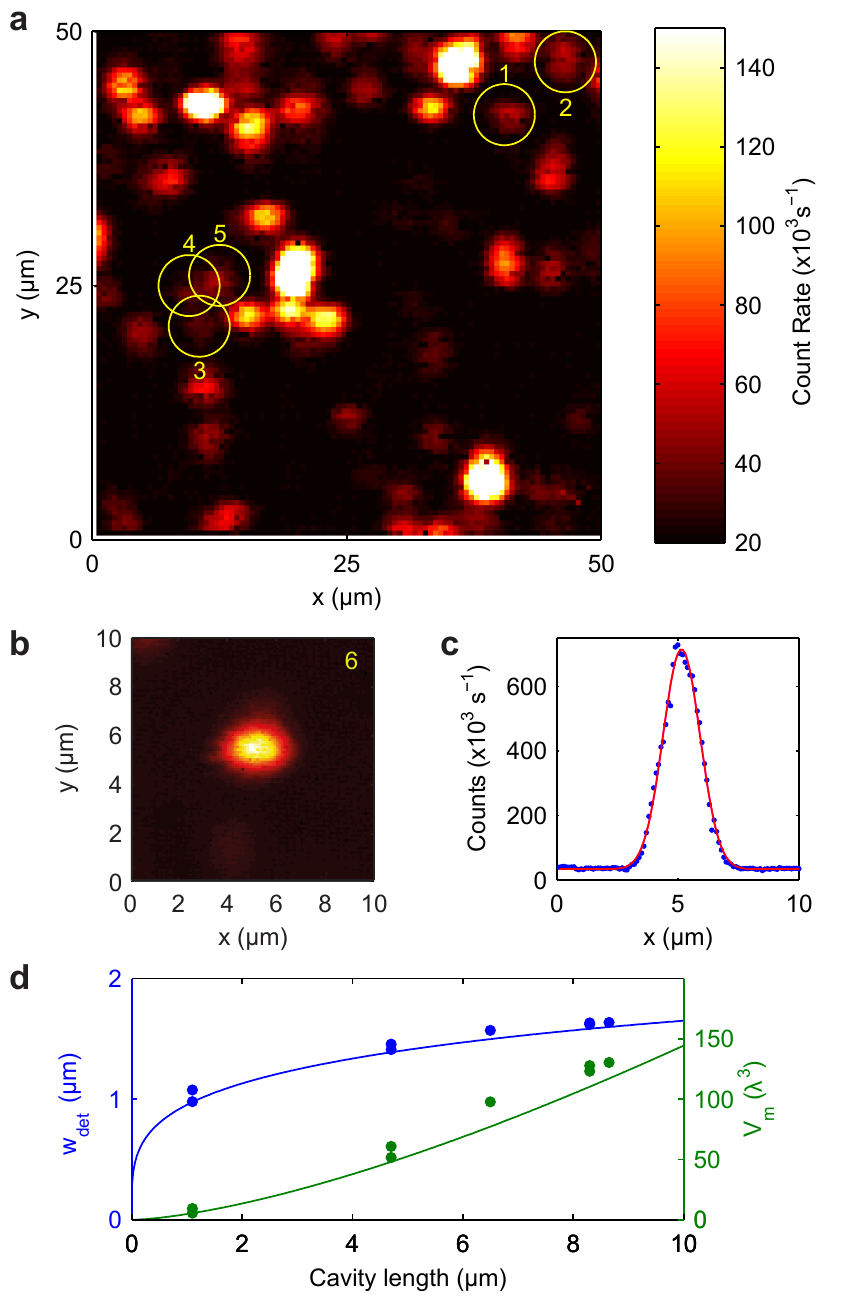}

\protect\caption{(a) Cavity-enhanced flourescence image showing several single emitters
(circles). The cavity length is stabilized on resonance with the excitation
light at a length of about $10\text{\,\textmu m}$ while scanned laterally.
(b) Cavity scan of a single NV center at $1.1\textrm{ \textmu m}$
cavity length. (c) Vertical cut through the scan in (b) together with
a Gaussian fit, yielding $w_{\textrm{det}}=1.1\,\text{\textmu m}$.
(d) Measured size of the point spread function (blue points) as a
function of the cavity length, together with the value calculated
from the radius of curvature of the fiber mirror (blue line). Mode
volume calculated from the measured $w_{\textrm{det}}$ (green points),
together with the expected value (green line).\label{fig:2_Scans}}
\end{figure}

We determine the mode volume of the cavity by measuring the optical
cavity length $d$ and the mode waist $w_{0}$, and use the expression
$V=\pi w_{0}^{2}d/4$. To determine $d$, we record the cavity transmission
under broadband illumination with a spectrometer, see Fig.~\ref{fig:1_Introduction}(f,g),
and evaluate the separation and location of the resonances. In this
way we prove that the shortest resonant cavity length $d=\lambda_{0}/2$
is actually reached without touching the planar mirror with the fiber
tip, i.e. full tunability is ensured even for the fundamental resonance
$q=1$. By scanning the cavity over a point-like object such as a
single NV center and collecting the fluorescence emitted into the
cavity (Fig.~\ref{fig:2_Scans}(b,c)), the mode waist can be inferred
from the observed size of the point spread function $w_{\textrm{det}}$
in a cavity-enhanced fluorescence image (see Appendix). We determine
a minimal $w_{\text{0}}=1.1\,$\textmu m at $d=\lambda_{0}/2$, such
that a minimal cavity mode volume $V=1.0\,\lambda_{0}^{3}$ ($0.34\,\mu\textrm{m}^{3}$)
is achieved.

\section{Cavity enhanced emission from Single NV centers}

We study a commercial nanodiamond sample (VanMoppes) with a size distribution between 100
and 200\,nm, where a reasonable fraction of the crystals contains
individual NV centers. The sample is directly spin coated or drop
cast onto the planar mirror. As a first step, we record a cavity-enhanced
fluorescence image of a large area of the sample by scanning the sample
mirror while the cavity length is stabilized to a mirror separation of
around $10\,\text{\textmu m}$, both the excitation light and the
NV emission spectrum being resonant with the cavity (see Appendix). Figure~\ref{fig:2_Scans}(a)
shows a scan area of $(50\,\text{\textmu m})^{2}$, where about 50
emitters are visible. The excitation power is 3.6\,mW at 532~nm.
The second-order correlation function $g^{(2)}(\tau)$, as well as
the saturation behavior for the five marked emitters is recorded.
A closer view on another, more isolated, single NV center (NV6) is
shown in Fig.~\ref{fig:2_Scans}(b,c), where the cavity length is
stabilized at $d=1.1$~\textmu m. This is the shortest mirror separation,
for which the cavity is simultaneously resonant for excitation and
fluorescence emission, such that significant Purcell enhancement is
present (see below). Since this yields the highest signal to background
ratio, we use it for all single-emitter measurements shown here. In
the measurement we observe a spatial resolution of 1.1\,\textmu m,
a high peak count rate ($K>6\cdot10^{5}\,\textrm{\ensuremath{\textrm{s}^{-1}}}$)
and a signal-to-background ratio >20 for the NV center showing clear
antibunching with $g^{(2)}(0)=0.21$. The non-zero values is primarily due 
to autofluorescence of the mirror coating. With the simultaneous excitation
and fluorescence enhancement and the emission into the well-collectable
cavity mode, the cavity enables enhanced count rates, and provides
spatial and spectral filtering at the same time. Cavity-enhanced scanning
fluorescence microscopy thus promises net signal improvement compared
to confocal microscopy. %

\begin{figure}
\includegraphics{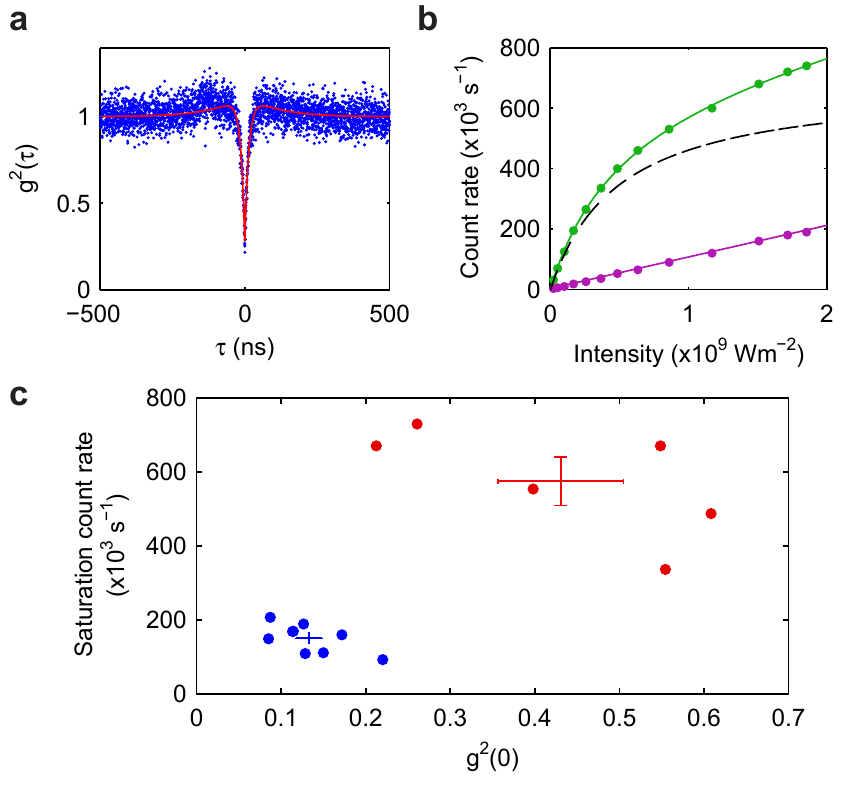}

\protect\caption{(a) Autocorrelation measurement for NV1 proving single emitter behavior
with $g^{(2)}(0)=0.27$ without background subtraction. (b) Saturation
measurement of the same NV center (green dots) together with a fit
(green line). Background fluorescence from the silver mirror (purple
dots) together with a linear fit (purple line). NV center fluorescence
subtracted by background (black dashed line). (c) Comparison of the
saturation count rates and single-photon purities for NV centers on
glass (blue dots) and inside the cavity at 1.1\,\textmu m cavity
length (red dots). An average enhancement by a factor 3.8 is found.\label{fig:3_PlaneConcaveSilverCavity}}
\end{figure}

As an example for the single-emitter fluorescence obtainable from
the cavity, we discuss the results for one NV center (NV1) in more
detail. We measure the $g^{(2)}$ function at low excitation power
and fit it with the function $g^{(2)}(\tau)=1+p(be^{-|\tau|/\tau_{2}}-(1+b)e^{-|\tau|/\tau_{1}}$),
which includes antibunching, bunching, and background, see Fig.~\ref{fig:3_PlaneConcaveSilverCavity}(a).
The fit value of $g^{(2)}(0)=0.27$ proves that it is a single emitter. Figure~\ref{fig:3_PlaneConcaveSilverCavity}(b) shows
a measurement of the intensity dependent fluorescence rate, which
is fitted well by a saturation part and a linear contribution accounting
for the background: $K=K_{\infty}I/(I_{sat}+I)+aI$ (see Appendix).
We find a saturation count rate $K_{\infty}=6.9\cdot10^{5}\,\textrm{\ensuremath{\textrm{s}^{-1}}}$,
a saturation intensity $I_{sat}=0.49\cdot10^{9}\,\textrm{W\ensuremath{m^{-2}}}$
and a linear background parameter $a=100\cdot10^{3}\,\textrm{s}^{-1}/10^{9}\,\textrm{W}\textrm{m}^{-2}$.
The intensity dependent background count rate is also recorded at
a location on the mirror without emitters. The observed slope is equal
to the value $a$ found for the emitter saturation, i.e. the diamond
crystal itself does not contribute notable background in this case.
The reported count rates are raw values as detected with avalanche
photo diodes (APDs). To obtain the number of photons collected by
the first lens, we account for the detection efficiency of the optical
setup of 43\% (see Appendix). From this we find that $1.6\times10^{6}$
photons per second are collected by the first lens.

To quantify the emission enhancement by the cavity, we study several
emitters in the cavity and compare the results with confocal measurements
(NA=0.75) of NV centers on a glass substrate.%
{} We record the $g^{(2)}$ function and the saturation count rate of
each emitter and compare the results by plotting the saturation count
rate versus $g^{(2)}(0)$ for cavity and free space emission, see
Fig.~\ref{fig:3_PlaneConcaveSilverCavity}(c). Taking background
fluorescence into account, we expect all NVs in the cavity with $g^{(2)}(0)\lesssim0.7$
being single emitters. The evaluation yields an average detected saturation
count rate per NV center of $K_{\infty}=5.7\cdot10^{5}\,s^{-1}$ inside
the cavity, in comparison with $1.5\cdot10^{5}\,s^{-1}$ on the glass
substrate. This corresponds to an average observed enhancement factor
of $3.8$. 

We estimate the theoretically expected enhancement as follows: The
photon rate coupled out of the cavity is given by $K_{c}=QE\cdot C_{\textrm{eff}}\gamma\eta_{c}$,
where $QE$ denotes the quantum efficiency and $\gamma$ the excited
state decay rate. For the experimental parameters used in the measurements
described above, applying the simple Purcell formula, we find $C_{\textrm{eff}}\approx0.12$,
for $V=5\lambda^{3}$, $Q_{c}=126$,\foreignlanguage{english}{ $Q_{\textrm{em}}=\frac{\lambda_{0}}{\Delta\lambda}\approx8$},
stemming from the center wavelength $\lambda_{0}=690\,\textrm{nm}$
and the FWHM $\Delta\lambda=90\,\textrm{nm}$ of the emission spectrum.
The fraction of photons leaving the cavity through the outcoupling
mirror is $\eta_{c}=\text{0.51}$, and the cavity mode is fully collected
by the objective. For the case of NV centers on a glass substrate,
the dipole radiation pattern is affected by the air-glass interface
\citep{Lukosz1977,Lukosz1979}. For the given NA we calculate a collection
efficiency of $\eta_{\Omega}=0.16$ and a photon collection rate $K=QE\cdot\gamma\eta_{\Omega}$.
The detection efficiency after the objective is assumed to be equal
to the cavity case. The total expected enhancement is then $K_{c}/K=C_{\textrm{eff}}\eta_{c}/\eta_{\Omega}=0.4$,
in contrast to the value found in the experiment. This shows that
the simplified treatment does not properly describe the situation.
In fact, the applied formula is derived in the limit where the cavity
mirrors subtend a negligible solid angle and where the dipole remains
far from any surface. For a more accurate treatment, we perform analytical
\citep{Lukosz1977,Lukosz1979} and finite difference time domain (FDTD)
simulations (Lumerical). We identify three aspects that influence
the spontaneous emission beyond the simple treatment: (i) The proximity
of the planar mirror surface with the sample leads to self-interference
of the dipole radiation over a large angular range as well as to some
amount of near-field coupling \citep{Lukosz1977,Lukosz1979}. A FDTD
simulation for an orientation-averaged dipole located in a nanodiamond
on a metal mirror predicts a lifetime reduction factor of 1.8 compared
to the situation without the mirror, as well as negligible nonradiative
decay (< 10\%) (see Appendix). (ii) The large solid angle subtended by the cavity
mirrors, having an NA close to unity, significantly affects the mode structure beyond the fundamental
cavity mode and consequently influences (e.g. inhibits) the decay
into modes other than the fundamental cavity mode \citep{hinds1991,Di2012}.
(iii) For suitable crystal size and smallest mirror separation, the
nanodiamond provides additional lateral mode confinement (see below).
All these aspects are captured by FDTD simulations of the emitter
inside the cavity, and we obtain an effective Purcell factor of $C_{\textrm{eff}}=1.4$
when comparing the cavity-coupled emission rate with free-space emission.
The predicted enhancement is then $K_{c}/K=C_{\textrm{eff}}\eta_{c}/\eta_{\Omega}=4.5$,
about twenty percent larger than determined experimentally. Expected
contributions to this deviation are a reduced coupling strength due
to non-ideal dipole orientation and a emitter location away from the
cavity field maximum.

\begin{figure*}
\includegraphics{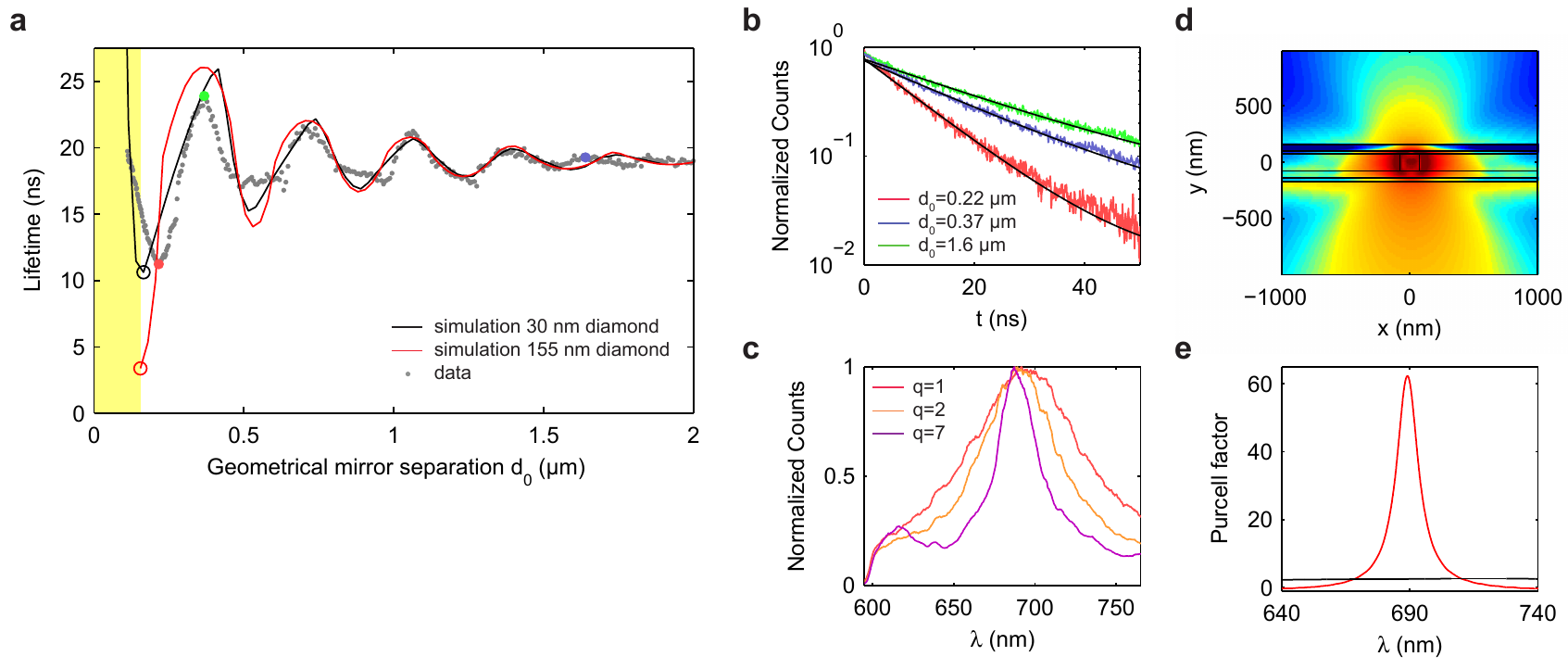}

\protect\caption{(a) Measured lifetime $\tau(d_{0})$ as a function of the mirror separation
(gray dots) together with FDTD simulations for a diamond cube of edge
length 30\,nm (black line) and 155\,nm (red line). The filled colored
dots indicate the positions at which the datasets shown in (b) were
taken. The yellow area shows the distance range where the fiber touches
the large diamond. (b) Lifetime traces for different cavity lengths
together with monoexponential fits (black), (red: $\tau=11.2\,\textrm{ns}$,
blue: $19.3\,\textrm{ns}$, green: $23.9\,\textrm{ns}$). (c) Cavity
emission spectra at resonances corresponding to longitudinal mode
orders $q=1,\,2,\,7$. (d) Simulation of the intensity distribution
of a dipole located at the center (logarithmic scale, red: high, blue:
low). The intensity is confined between the two silver mirrors (horizontal
black lines, including the spacer layers) and localized to the nanocrystal
(black square). Outcoupling occurs predominantly through the thinner
bottom mirror. (e) Simulated Purcell factor as a function of the wavelength
for a 155\,nm (red, corresponding to open circle in (a)) and a 30\,nm
(black) diamond. \label{fig:4_PlanePlaneCavity}}
\end{figure*}

\section{Lifetime modification}

A central aspect of the Purcell enhancement is the influence of the
cavity on the excited state lifetime. To demonstrate this effect we
perform time-correlated single photon counting under pulsed excitation at different mirror separations. In measurements with single emitters we observe significant
lifetime variation. However, we find that the lifetime is also affected
by varying amounts of background, related to the distance dependent
cavity coupling efficiency of the excitation light. To unambiguously
demonstrate the effect of the cavity on the lifetime, we therefore
study ensembles of NV centers, where background plays a negligible
role. We use nanodiamonds with a diameter distribution peaking between
100 and 150\,nm that were irradiated with He ions to achieve a high
concentration of NV centers ($\sim10^{2}$ per crystal) \citep{Mohan2010,Yu2005}.
In these measurements, the used cavity fiber has a planar endfacet
whose edges are mechanically polished off to allow for smallest mirror
separations. This simpler geometry stemming from an earlier approach
is technically easy to realize and facilitates direct comparison with
FDTD simulations. The fiber has an identical coating as in the previous
case, while the large mirror has higher transmission due to a thicker
glass capping (60\,nm), resulting in $\mathcal{F}=28$ at $700\,\textrm{nm}$.

We measure the fluorescence lifetime of the NV center ensemble for
different mirror separations by time-correlated single photon counting \citep{Chizhik2011}, where we tune the
mirror separation with a piezo actuator and calibrate the optical
cavity length $d$ as well as the geometrical mirror separation $d_{0}$
from fluorescence spectra \footnote{In the measurement shown in Fig. 4(e), we observe a deviation of the
actual cavity length from the expected value below $d\sim500$~nm,
most probably due to the cavity fiber touching the large mirror with
one of its edges. We have adjusted the distance calibration below
500~nm to restrict the cavity lengths to $d>100$~nm.}. Each decay trace is fitted with a monoexponential decay, from which
we obtain the lifetime $\tau(d_{0})$ \footnote{We note that in general, a multi-exponential decay is expected due
to the position and dipole orientation distribution of the colour
centers in the nanodiamond. Fitting the data with stretched exponentials
\citep{Berberan-Santos2005} yields a larger lifetime modulation,
such that the evaluation shown is a conservative estimate.}. Figure~\ref{fig:4_PlanePlaneCavity}(a) shows an example measurement.
Figure~\ref{fig:4_PlanePlaneCavity}(b) presents individual traces
for the mirror separations indicated in (a). At the same time, the
fluorescence is spectrally filtered by the cavity, yielding a resonance
with a linewidth depending inversely on the mirror separation, see
Fig.~\ref{fig:4_PlanePlaneCavity}(c). For cavity lengths below 2\,\textmu m,
the lifetime is modulated noticeably. Whenever the cavity is resonant
(off-resonant) with the emission spectrum, the lifetime is reduced
(increased) due to the variation of the local density of states. The
shortest (longest) detected lifetime of $\tau_{c}=11.2\,\textrm{ns}$
($23.7\,\textrm{ns}$) corresponds to a lifetime reduction (enlargement)
of 40\,\% (25\,\%) compared to the lifetime of $\tau_{m}=18.9\,\textrm{ns}$
obtained when the cavity fiber is not present. We estimate the effect
of the sample mirror alone from a statistical comparison of lifetimes
both on the mirror and on a glass substrate (reduction by a factor
of $1.3\pm0.3$), as well as from FDTD simulations averaged over dipole
orientations (reduction factor $1.8$), suggesting a free-space lifetime
of up to $\tau_{0}=34\,\textrm{ns}$. With this, we obtain an upper
bound for the effective Purcell factor of $C_{\textrm{eff}}=\frac{\tau_{0}}{\tau_{c}}-1=2.0$.
We have studied several nanocrystals in a similar manner and observe
a shortest lifetime of $\tau_{c}=6.7\,\textrm{ns}$ with comparable
relative lifetime changes. At the same time, the integrated fluorescence
increases by up to a factor of 40 from $d=2\,\lyxmathsym{\textmu}\textrm{m}$
to $d=\lambda/2$. This is due to the fact that the cavity has a negligible effect
for large mirror separations and only the emission transmitted through the planar mirror ($T_1=0.11$) is collected with a small collection efficiency $\eta_{\Omega}=0.16$. In contrast, for short cavities, the Purcell effect leads to a dominant emission into the cavity mode, a fraction $\eta_{c}=0.56$ of which is coupled out to the detector side and entirely collected by the objective (see Appendix). 

We compare our results to three dimensional FDTD simulations. The
diamond is modeled as a cube with a refractive index of 2.4 and an
edge length of 155\,nm. In addition, we study a 30\,nm crystal for
comparison. The individual mirror layers are implemented with parameters
as given above. A dipole source peaking at a wavelength of 690\,nm
with a spectral width of 100\,nm is placed at the center of the cube.
We perform simulations with a dipole orientation parallel and normal
to the plane of the mirrors for different mirror separations. Figure~\ref{fig:4_PlanePlaneCavity}(d)
shows a cross section of the geometry with a dipole oriented along
the $x-$axis located in the center. A strong confinement of the intensity
to the crystal is visible (note the logarithmic color scale), as well
as the directional outcoupling through the bottom mirror. The outcoupled
mode has an NA of 0.26 (0.38 after refraction at an air-glass interface),
which can be easily collected by an objective. For each cavity length,
the simulation evaluates the Purcell factor as a function of the wavelength
$C(\lambda)$. Figure~\ref{fig:4_PlanePlaneCavity}(e) shows an example
for the 155\,nm crystal at $d=\lambda/2$, yielding a peak Purcell
factor of $C=63$ compared to the nanocrystal in vacuum. To enable
a comparison with the experimental data, we normalize $C(\lambda)$
to the Purcell factor $C_{m}(\lambda)$ obtained for simulations where
the fiber mirror is removed, and average over the NV emission spectrum
$S(\lambda)$. The spectrum is modeled as a Gaussian distribution
of 110~nm $1/e\textrm{-full-width}$, centered at 690\,nm. The effective
enhancement factor is then determined as $\tau_{c}/\tau_{m}=\int_{_{-\infty}}^{^{\infty}}C(\lambda)C_{m}(\lambda)^{-1}S(\lambda)d\lambda/\int_{_{-\infty}}^{^{\infty}}S(\lambda)d\lambda$.
The FDTD simulations for the parallel dipole are in good agreement
with the measurements. Contributions of the normal dipole component
play a negligible role, since this orientation is only weakly excited
and coupled to the cavity. For the 155\,nm diamond, a maximal effective
Purcell factor of $C_{\textrm{eff}}=11$ is predicted when the fiber
touches the diamond. In contrast, the 30~nm diamond yields a much
smaller Purcell factor, and the enhancement spectrum $C(\lambda)$
does not show a clear resonance (see Fig.~\ref{fig:4_PlanePlaneCavity}(e)).

\section{Waveguide effect}

We explain the large Purcell factor by additional mode confinement
of suitably sized nanocrystals. Some intuition about the confinement
effect can be drawn from modeling the crystal as a cylindrical waveguide
\citep{Babinec2010} with refractive index $n_{d}=2.4$ and solving
the Helmholtz equation for the electric field $(\nabla^{2}+k^{2})E=0$
in one dimension. Here, $k=2\pi n_{r}/\lambda$, with $n_{r}=n_{d}$
for $|r|\leq b$ and 1 for $|r|>b$, $r$ is the transverse coordinate,
and $b$ the waveguide radius. We find that the propagating waveguide
mode shows strongest confinement around $b=70$~nm, yielding a minimal
effective mode radius ($1/e^{2}$) of $w_{0}=160$\,nm. A segment
of such a nanowire can now be considered to be introduced into a planar
Fabry-Perot cavity with mirror separation $d=\lambda/2n_{\mathrm{eff}}$,
where \foreignlanguage{english}{$n_{\mathrm{eff}}$} is the effective
refractive index experienced by the propagating mode (we find \foreignlanguage{english}{$n_{\mathrm{eff}}=1.88$}
for $b=70$~nm). The mode volume of such a cavity would amount to
$V=\pi w_{0}^{2}\lambda/8n_{\mathrm{eff}}=0.07\,(\lambda/n_{\mathrm{eff}})^{3}$,
substantially smaller than what is achievable by conventional curved-mirror
Fabry-Perot cavities. Together with an effective quality factor $Q_{\textrm{eff}}=8$,
this yields $C_{\textrm{eff}}=8$, in reasonable agreement with the
FDTD simulation. From the more reliable FDTD simulations we find an
optimal crystal size around 155\,nm for a cubic shape. A $\pm10$\,\%
size variation does not significantly affect the maximal $C$, while
the enhancement rapidly diminishes for crystals < 140\,nm and no
additional enhancement is obtained for the 30\,nm crystal.

The measurement also suggests the presence of a contribution of the
waveguide effect. Experimental imperfections, averaging over dipole
orientation, and finite quantum efficiency are expected to lead to
a reduced lifetime modification. This is evident most clearly e.g.
at the lifetime minima corresponding to the resonances $q=2$ and
$q=3$, where the lifetime reduction in the data remains smaller than
in the simulation for both crystal sizes \footnote{We note that a small amount of parasitic lifetime modulation seems
to be also present, again most probably due to the varying amount
of excitation light coupled into the cavity and the corresponding
variation of background contribution. It can be seen at large mirror
separation (e.g. for $d>1.3\,\textrm{\ensuremath{\mu\textrm{m}}}$)
with a periodicity of 270~nm.}. However, for $q=1$, where the waveguide effect is expected to be
present, the experimental variation is as large as the simulated ideal
variation for the 30\,nm crystal. Consequently, the Purcell factor
in the experiment needs to be larger than in the simulation to compensate
for the imperfections. The increase of the measured lifetime towards
smallest $d$ indicates a crystal size below 140\,nm, which could
in part explain the smaller effect.

\section{Conclusion}

We have demonstrated that improved laser machining enables the realization
of stable Fabry-Perot cavities with mode volumes down to $1\,\lambda^{3}$,
while maintaining full tunability. Multiple emitters can be investigated
with a single cavity, and efficient single photon extraction is possible
even for broadband emitters. The Purcell enhancement together with
high outcoupling efficiency allows net count rates exceeding free-space
collection. Furthermore, we have shown with simulations that suitably
chosen nanodiamonds can provide exceptional mode confinement. Predicted
ideal Purcell factors up to 63 could be fully exploited with narrow-band
emitters such as the SiV center in diamond, and effective Purcell
factors for NV centers up to 11 are expected. While our experiments
already indicate a contribution due to this effect, we expect that
more controllably fabricated diamond nanostructures \citep{Arend2016,Maletinsky2012,Momenzadeh2015}
can unfold the full potential of this approach and pave the way for
ultra-bright single photon sources and superior readout of single
spins under ambient conditions.

\section*{Acknowledgements}

We thank Aniket Agrawal and Tolga Bagci for contributions to the experiment,
as well as Philipp Altpeter for assistance in the cleanroom. Fruitful
discussions with Christoph Becher, Jason Smith, and J\"org Wrachtrup are acknowledged.
The work has been funded by the European Union 7th framework Program
under grant agreement no. 61807 (www.fp7wasps.org) and the DFG Cluster
of Excellence NIM. T. W. Hänsch acknowledges funding from the Max-Planck
Foundation.

\section*{Appendix A: Cavity characterization and setup}

\paragraph*{Mirror coating}

The fiber tips are coated with a $60\,\textrm{nm}$ silver film and
finished with a $20\,\textrm{nm}$ glass capping to prevent oxidation.
Consistent with simulations \citep{Furman1992} we measure a reflectivity
of $R_{2}=(92\pm2)\%$ at $532\,\textrm{nm}$. The macroscopic planar
mirror is prepared on a low autofluorescence glass substrate with
a $33\,\textrm{nm}$ thick silver layer and an analogous glass capping.
At a wavelength of $532\,\textrm{nm}$ we measure a mirror transmission
of $T_{1}=15\%$. A simulation yields the transmission and absorption
loss of the two mirrors at 700\,nm: $T_{1}=8\%$, $T_{2}=0.8\%$
and $A_{1}=4\%$, $A_{2}=3\%$. With atomic force microscopy measurements
we determine the surface roughness of the silver mirros to be $<5\,\textrm{nm rms}$,
such that scattering loss plays a negligible role ($<0.5\%$). The
measured finesse agrees well with the transmission and loss values,
and we calculate the expected fraction of photons leaving the cavity
through the macroscopic mirror to be $\eta_{c}=T_{1}/(T_{1}+T_{2}+A_{1}+A_{2})=0.51$.
Despite the lossy character of the metal coatings, a high outcoupling
efficiency is achieved. For the plane-plane cavity, the glass capping
layer of the macroscopic mirror is increased to 60\,nm, resulting
in an increased transmission, $T_{1}=24\%$ at 532\,nm and $11\%$ at 700~nm, respectively. The outcoupling efficicincy amounts to $\eta_{c}=0.56$ at 700~nm.

\paragraph*{Mode volume}

The mode volume of a plano-concave cavity is given by $V=\pi w_{0}^{2}d/4$
, with the mode waist $w_{0}^{2}=\lambda/\pi\cdot\sqrt{r_{c}d-d^{2}}$
and the optical cavity length $d=\frac{\lambda}{2}\left[q+\frac{\zeta}{\pi}\right]\approx q\frac{\lambda}{2}$
for $r_{c}\gg d$. Here, $q$ is the longitudinal mode order, and
$\zeta=\arccos\sqrt{1-d/r_{c}}$ is the Gouy phase. The finite conductivity
of the silver mirror and the capping layer leads to some field penetration
into the mirror, such that the geometrical mirror separation $d_{0}=\frac{\lambda}{2}\left[q+\frac{\zeta-\phi}{\pi}\right]$
is smaller than $d$. This is accounted for by the average deviation
of the reflection phase from $\pi$, \foreignlanguage{english}{$\phi=\pi-(\phi_{1}+\phi_{2})/2$}.
At $\lambda=700\,\textrm{nm}$, $\phi_{1}\approx\phi_{2}=0.72\thinspace\pi$
for our mirrors \citep{Becker1997}, and we calculate an air gap of
$d_{0}=260$~nm for $q=1$. This value reduces further when a nanodiamond
of significant size is placed in the cavity mode due to the effective
refractive index change.

\paragraph*{Point spread function}

In cavity-enhanced fluorescence scans, the observed point spread function
is the product of the cavity mode for emission ($w_{\textrm{0}}$)
and excitation ($w_{\textrm{e}}=\sqrt{\lambda_{\textrm{e}}/\lambda}w_{\textrm{0}}$).
The size of the point spread function is then $w_{\textrm{det}}=w_{\textrm{e}}w_{\textrm{0}}/\sqrt{w_{\textrm{e}}^{2}+w_{\textrm{0}}^{2}}$,
which can be used to determine $w_{0}$. The experimental values agree
well with a calculation for different cavity lengths when using the
measured radius of curvature of the fiber mirror, see (Fig.~\ref{fig:2_Scans}(d)).

\paragraph*{Setup}

The cavity is embedded into a confocal microscope setup, similar to
the one described in \citep{Kaupp2013}. Excitation is performed with
either a cw laser at $532\,\textrm{nm}$ or a band-pass filtered supercontinuum
source ($\sim50\thinspace\textrm{ps}$ pulse length, 20\,MHz repetition
rate) through a cover slip corrected microscope objective (NA=0.75)
through the planar mirror. Alternatively, the excitation can be performed
through the cavity fiber. For pulsed excitation we use average excitation
powers ranging from a couple of ten to a few hundred \textmu W. The
planar mirror is mounted on a three-axis slip-stick nanopositioner
allowing to scan the mirror for suitable emitters. The fiber is mounted
on a stacked piezo actuator which allows for precise tuning and stabilization
of the cavity length. The latter is performed by using the green excitation 
light transmitted through the cavity to generate a cavity length-dependent 
feedback signal, which is fed to the stacked piezo controlling the fiber position.

\paragraph*{Excitation intensity}

For the saturation measurements and autocorrelation measurements of
single emitters in the cavity (Fig.~\ref{fig:3_PlaneConcaveSilverCavity})
we excite through the cavity fiber with 532\,nm light. We calculate
the excitation intensity $I$ inside the cavity from the measured
transmitted power $P_{t}$, the outcoupling mirror transmission $T_{1}$,
and the evaluated excitation mode waist $w_{e}$ according to $I=\frac{8P_{t}}{\pi w_{e}^{2}T_{1}}$.

\paragraph*{Detection efficiency}

We infer the detection efficiency of the setup by coupling a laser
at $690\,\textrm{nm}$ into the cavity fiber with the cavity on resonance
and measure the transmission of the mode coupled out of the cavity
through the optics up to the APDs. A fraction of $67\%$ of the light
reaches the detectors at a cavity length of 1\,\textmu m. Together
with the quantum efficiency of the APDs of $65\%$, a detection efficiency
of 43\% at 690\,nm is achieved.

\paragraph*{Nonradiative decay}
To study the onset of nonradiative decay, we simulated the quantum efficiency 
of a dipole in a diamond cube (edge length 30\,nm) on a silver mirror 
(layer thickness 33\,nm) with an intermediate glass spacer layer of variable thickness. 
For glass layers larger than 30\,nm, the quantum efficiency exceeds 90\% for dipoles parallel 
to the mirror surface. Since the applied spacer layers have a thickness of 60\,nm, 
we expect the amount of nonradiative decay to be less than 10\%.

%

\end{document}